\documentclass{emulateapj} \usepackage{apjfonts}

\newcommand{\Ha}{$\rm{H} \alpha$}

\newcommand{\etal}{et~al.}
\newcommand{\PVdblt}{{\rm P}\kern 0.1em{\sc v}~$\lambda\lambda 1117, 1128$}
\newcommand{\CaIIdblt}{{\rm Ca}\kern 0.1em{\sc ii}~$\lambda\lambda 3934, 3969$}
\newcommand{\AlIIIdblt}{{\rm Al}\kern 0.1em{\sc iv}~$\lambda\lambda 1855, 1863$}
\newcommand{\CIVdblt}{{\rm C}\kern 0.1em{\sc iv}~$\lambda\lambda 1548, 1550$}
\newcommand{\MgIIdblt}{{\rm Mg}\kern 0.1em{\sc ii}~$\lambda\lambda 2796, 2803$}
\newcommand{\NVdblt}{{\rm N}\kern 0.1em{\sc v}~$\lambda\lambda 1238, 1242$}  
\newcommand{\SVIdblt}{{\rm S}\kern 0.1em{\sc vi}~$\lambda\lambda 933, 944$} 
\newcommand{\OVIdblt}{{\rm O}\kern 0.1em{\sc vi}~$\lambda\lambda 1031, 1037$} 
\newcommand{\SiIIdblt}{{\rm Si}\kern 0.1em{\sc ii}~$\lambda\lambda 1190, 1193$} 
\newcommand{\SiIVdblt}{{\rm Si}\kern 0.1em{\sc iv}~$\lambda\lambda 1393, 1402$} 
\newcommand{\PV}{\hbox{{\rm P}\kern 0.1em{\sc v}}}
\newcommand{\AlI}{\hbox{{\rm Al}\kern 0.1em{\sc i}}}
\newcommand{\AlII}{\hbox{{\rm Al}\kern 0.1em{\sc ii}}}
\newcommand{\AlIII}{{\hbox{\rm Al}\kern 0.1em{\sc iii}}}
\newcommand{\CaII}{\hbox{{\rm Ca}\kern 0.1em{\sc ii}}}
\newcommand{\CII}{\hbox{{\rm C}\kern 0.1em{\sc ii}}}
\newcommand{\CIIe}{\hbox{{\rm C$^{\ast}$}\kern 0.1em{\sc ii}}}
\newcommand{\CIII}{\hbox{{\rm C}\kern 0.1em{\sc iii}}}
\newcommand{\CIV}{\hbox{{\rm C}\kern 0.1em{\sc iv}}}
\newcommand{\CV}{\hbox{{\rm C}\kern 0.1em{\sc v}}}
\newcommand{\HI}{\hbox{{\rm H}\kern 0.1em{\sc i}}}
\newcommand{\HII}{\hbox{{\rm H}\kern 0.1em{\sc ii}}}
\newcommand{\Lya}{\hbox{{\rm Ly}\kern 0.1em$\alpha$}}
\newcommand{\Lyb}{\hbox{{\rm Ly}\kern 0.1em$\beta$}}
\newcommand{\Lyg}{\hbox{{\rm Ly}\kern 0.1em$\gamma$}}
\newcommand{\Lyd}{\hbox{{\rm Ly}\kern 0.1em$\delta$}}
\newcommand{\Lye}{\hbox{{\rm Ly}\kern 0.1em$\epsilon$}}
\newcommand{\Lyphi}{\hbox{{\rm Ly}\kern 0.1em$\phi$}}
\newcommand{\Lyfive}{\hbox{{\rm Ly}\kern 0.1em$5$}}
\newcommand{\Lysix}{\hbox{{\rm Ly}\kern 0.1em$6$}}
\newcommand{\Lyseven}{\hbox{{\rm Ly}\kern 0.1em$7$}}
\newcommand{\Lyeight}{\hbox{{\rm Ly}\kern 0.1em$8$}}
\newcommand{\Lynine}{\hbox{{\rm Ly}\kern 0.1em$9$}}
\newcommand{\Lyten}{\hbox{{\rm Ly}\kern 0.1em$10$}}
\newcommand{\Lyeleven}{\hbox{{\rm Ly}\kern 0.1em$11$}}
\newcommand{\HeI}{\hbox{{\rm He}\kern 0.1em{\sc i}}}
\newcommand{\HeII}{\hbox{{\rm He}\kern 0.1em{\sc ii}}}
\newcommand{\FeI}{\hbox{{\rm Fe}\kern 0.1em{\sc i}}}
\newcommand{\FeII}{\hbox{{\rm Fe}\kern 0.1em{\sc ii}}}
\newcommand{\FeIII}{\hbox{{\rm Fe}\kern 0.1em{\sc iii}}}
\newcommand{\MnII}{\hbox{{\rm Mn}\kern 0.1em{\sc ii}}}
\newcommand{\MgI}{\hbox{{\rm Mg}\kern 0.1em{\sc i}}}
\newcommand{\MgIb}{\hbox{{\rm Mg}\kern 0.1em{\sc i}}\kern 0.05em{\rm b}}
\newcommand{\MgII}{\hbox{{\rm Mg}\kern 0.1em{\sc ii}}}
\newcommand{\MgIII}{\hbox{{\rm Mg}\kern 0.1em{\sc iii}}}
\newcommand{\NI}{\hbox{{\rm N}\kern 0.1em{\sc i}}}
\newcommand{\NII}{\hbox{{\rm N}\kern 0.1em{\sc ii}}}
\newcommand{\NIII}{\hbox{{\rm N}\kern 0.1em{\sc iii}}}
\newcommand{\NV}{\hbox{{\rm N}\kern 0.1em{\sc v}}}
\newcommand{\OVI}{\hbox{{\rm O}\kern 0.1em{\sc vi}}}
\newcommand{\OI}{\hbox{{\rm O}\kern 0.1em{\sc i}}}
\newcommand{\OII}{\hbox{[{\rm O}\kern 0.1em{\sc ii}]}}
\newcommand{\OIII}{\hbox{[{\rm O}\kern 0.1em{\sc iii}]}}
\newcommand{\OIV}{\hbox{{\rm O}\kern 0.1em{\sc iv}]}}
\newcommand{\SI}{{\rm S}\kern 0.1em{\sc i}}
\newcommand{\SIV}{{\rm S}\kern 0.1em{\sc iv}}
\newcommand{\SVI}{{\rm S}\kern 0.1em{\sc vi}}
\newcommand{\SiI}{\hbox{{\rm Si}\kern 0.1em{\sc i}}}
\newcommand{\SiII}{\hbox{{\rm Si}\kern 0.1em{\sc ii}}}
\newcommand{\SiIII}{\hbox{{\rm Si}\kern 0.1em{\sc iii}}}
\newcommand{\SiIV}{\hbox{{\rm Si}\kern 0.1em{\sc iv}}}
\newcommand{\SII}{\hbox{{\rm S}\kern 0.1em{\sc ii}}}
\newcommand{\SIII}{\hbox{{\rm S}\kern 0.1em{\sc iii}}}
\newcommand{\NaI}{\hbox{{\rm Na}\kern 0.1em{\sc i}}}
\newcommand{\NaID}{\hbox{{\rm Na}\kern 0.1em{\sc i}}\kern 0.05em{\rm D}}
\newcommand{\TiII}{\hbox{{\rm Ti}\kern 0.1em{\sc ii}}}
\newcommand{\kms}{\hbox{~km~s$^{-1}$}}

\newcommand{\Mpyr}{\hbox{~M$_{\odot}$yr$^{-1}$}}
\newcommand{\kpc}{\hbox{~kpc}}
\newcommand{\A}{\hbox{~\AA}}

\newcommand{\ack}{We would like to thank the referee for his/her
  thorough read of the manuscript. We thank Sowgat Mazuhid for
  providing the {\it HST}/COS spectrum. This work was partly supported
  by a Marie Curie International Outgoing Fellowship
  (PIOF-GA-2009-236012), by a Marie Curie International Career
  Integration Grant (PCIG11-GA-2012-321702) and by NSF grant
  AST-1109288.  Data was obtained at the W.M. Keck Observatory, which
  is operated as a scientific partnership among the California
  Institute of Technology, the University of California and the
  National Aeronautics and Space Administration. The Observatory was
  made possible by the generous financial support of the W.M. Keck
  Foundation. Additional observations were obtained with the APO 3.5-m
  telescope, which is owned and operated by the Astrophysical Research
  Consortium. Observations were also made with the NASA/ESA HST or
  obtained from the data archive at the Space Telescope Institute.
  This research used the facilities of the Canadian Astronomy Data
  Centre operated by the National Research Council of Canada with the
  support of the Canadian Space Agency.}
\slugcomment{Submitted V2 April 30, 2014} 
\shorttitle{\sc New perspective on outflows}
\shortauthors{\sc Kacprzak et~al.}

\begin{document}


\title{New perspective on galaxy outflows from the first detection of
  both intrinsic and traverse metal-line absorption}


\author{\sc
Glenn G. Kacprzak\altaffilmark{1,2},
Crystal L. Martin\altaffilmark{3},
Nicolas Bouch{\'e}\altaffilmark{4,5},
Christopher W. Churchill\altaffilmark{6},
Jeff Cooke\altaffilmark{1}, \\
Audrey LeReun\altaffilmark{4,5},
Ilane Schroetter\altaffilmark{4,5}, 
Stephanie H. Ho\altaffilmark{3},
Elizabeth Klimek\altaffilmark{6}
}

\altaffiltext{1}{Swinburne University of Technology, Victoria 3122,
Australia {\tt gkacprzak@astro.swin.edu.au}}
\altaffiltext{2}{Australian Research Council Super Science Fellow}
\altaffiltext{3}{Physics Department, University of California, Santa Barbara, CA 93106, USA}
\altaffiltext{4}{CNRS, Institut de Recherche en Astrophysique et Plan{\'e}tologie [IRAP] de Toulouse, 14 Avenue E. Belin, F-31400 Toulouse, France}
\altaffiltext{5}{Universit{\'e} Paul Sabatier de Toulouse, UPS-OMP, IRAP, F-31400 Toulouse, France}
\altaffiltext{6}{New Mexico State University, Las Cruces, NM 88003}

\begin{abstract}

  We present the first observation of a galaxy ($z=0.2$) that exhibits
  metal-line absorption back-illuminated by the galaxy
  (``down-the-barrel'') {\it and} transversely by a background quasar
  at a projected distance of 58{\kpc}. Both absorption systems, traced
  by {\MgII}, are blueshifted relative to the galaxy systemic
  velocity.  The quasar sight-line, which resides almost directly
  along the projected minor axis of the galaxy, probes {\MgI} and
  {\MgII} absorption obtained from Keck/LRIS and {\Lya}, {\SiII} and
  {\SiIII} absorption obtained from {\it HST}/COS. For the first time,
  we combine two independent models used to quantify the outflow
  properties for down-the-barrel and transverse absorption.  We find
  that the modeled down-the-barrel deprojected outflow velocities
  range between $V_{dtb}=45-255${\kms}. The transverse bi-conical
  outflow model, assuming constant-velocity flows perpendicular to the
  disk, requires wind velocities $V_{outflow}=40-80${\kms} to
  reproduce the transverse {\MgII} absorption kinematics, which is
  consistent with the range of $V_{dtb}$. The galaxy has a
  metallicity, derived from H$\alpha$ and {\NII}, of
  [O/H]=$-0.21$$\pm$0.08, whereas the transverse absorption has $[{\rm
    X/H}]=-1.12\pm0.02$.  The galaxy star-formation rate is
  constrained between 4.6--15{\Mpyr} while the estimated outflow rate
  ranges between 1.6--4.2{\Mpyr} and yields a wind loading factor
  ranging between $0.1-0.9$.  The galaxy and gas metallicities, the
  galaxy--quasar sight-line geometry, and the down-the-barrel and
  transverse modeled outflow velocities collectively suggest that the
  transverse gas originates from ongoing outflowing material from the
  galaxy. The $\sim$1~dex decrease in metallicity from the base of the
  outflow to the outer halo suggests metal dilution of the gas by the
  time it reached 58{\kpc}.

\end{abstract}



\keywords{galaxies: halos --- galaxies: intergalactic medium ---
  quasars: absorption lines}

%

\section{Introduction}
\label{sec:intro}

Galactic-scale outflows appear to be quite common amongst star-forming
galaxies and these outflows are thought to contribute significantly to
the metal enrichment of the universe
\citep[e.g.,][]{oppenheimer10}. Cool gas tracers such as the
{\MgIIdblt} doublet are commonly used to trace and measure outflow
properties.

Star-forming galaxies show intrinsic blueshifted outflowing {\MgII}
gas with velocities of 100--1000{\kms}
\citep[e.g.,][]{tremonti07,weiner09,martin09,coil11,martin12,rubin13}
and occasionally show redshifted infalling gas with velocities of
50--200{\kms} \citep[e.g.,][]{rubin12,martin12,rubin13}. These studies
of systems back-illuminated by the galaxy (``down-the-barrel'') have
shown that outflowing gas velocities weakly correlate with galaxy
star-formation rates (SFR) and specific star-formation rates
\citep{martin05,weiner09,martin12,rubin13}.  The outflow velocities
also correlate with galaxy inclination, with higher velocities
occurring for face-on galaxies \citep{kornei12,bordoloi13}. Although
these observations constrain wind velocities and how they relate to
their galaxies, they do not constrain their extent or the mass
ejection rates.

Background quasars have been used to probe the ``transverse''
absorption associated with outflows, which is supported by a strong
azimuthal dependence of {\MgII} absorption around galaxies
\citep{bouche12,kacprzak12a,bordoloi14}. \citet{bouche12} have also
shown that transverse absorption detected along the projected minor
axes of galaxies is kinematically consistent with simple bi-conical
outflows, leading to constraints on mass outflow rates. The azimuthal
dependence, together with an inclination-dependence in down-the-barrel
studies, points towards a picture where outflows are well collimated
with half-opening angles of $\sim$45 degrees
\citep{bouche12,kacprzak12a,martin12,bordoloi13,bordoloi14}.


\begin{figure*}
\begin{center}
\includegraphics[angle=0,scale=0.40]{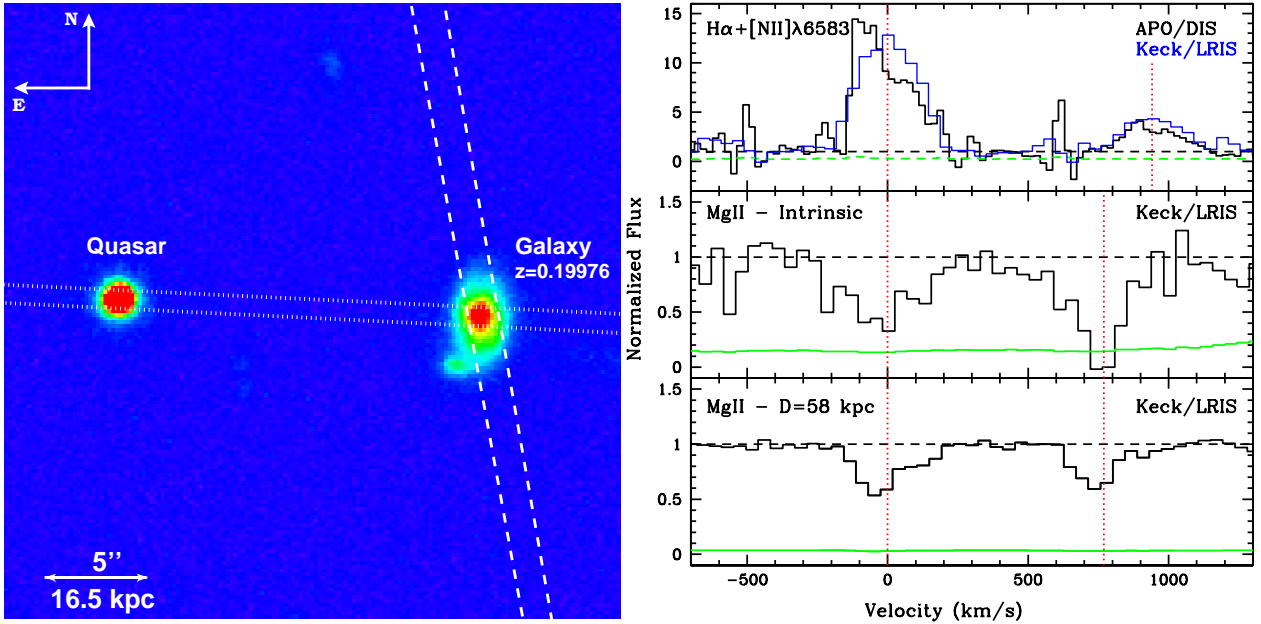}
\caption[angle=0]{ (left) 30$''$$\times$30$''$ $i$-band MegaCam/CFHT
  image of the background quasar and the z=0.19976 {\MgII} absorbing
  galaxy at a projected distance of 58{\kpc}. The dashed-line shows
  the DIS/APO slit placed along the galaxy major-axis and the
  dotted-line shows the LRIS/Keck slit placed along the
  minor-axis. --- (right, top) {\Ha}+{[\NII]} emission along the major
  axis (DIS/APO, black) and along the minor axis (Keck/LRIS, blue). In
  all panels, red vertical-lines define the {\Ha} centroid and galaxy
  systemic velocity. (right, middle) LRIS/Keck spectrum of the
  {\MgIIdblt} absorption ``down-the-barrel'' of the galaxy, which is
  blueshifted relative to {\Ha}. (right, bottom) {\MgIIdblt}
  absorption along the quasar sight-line, which is also blueshifted.}
\label{fig:image}
\end{center}
\end{figure*}

Simulations suggest that the metallicity of transverse absorption can
also be used to constrain its origins \citep[e.g.,][]{shen12}.  Low
metallicity gas near galaxies has been assumed to be infalling
\citep[e.g., ][]{kacprzak12b,bouche13}, while high metallicity gas
near galaxies has also been assumed to be outflowing
\citep{peroux11,stocke13}.

Despite the detailed and large studies of transverse and
down-the-barrel absorption systems, the two have yet to be directly
connected.  We have begun a Keck-SDSS QSO-galaxy pair survey,
expanding on the pioneering work of \citet{barton09}, to study the CGM
using {\MgII} absorption around $z=0.1-0.3$ galaxies.  Here we present
our initial findings of the first observation that directly connects
down-the-barrel outflows with the transverse gas observed 58{\kpc}
from a $z=0.2$ galaxy.  In \S~\ref{sec:data} we describe the data and
our analysis.  In \S~\ref{sec:results}, we present kinematic models of
the down-the-barrel and transverse absorption profiles and show that
they are both consistent with wind models. We use {\it HST}/COS
observations to compute the metallicity of the transverse gas. We end
with a discussion and concluding remarks in
\S~\ref{sec:discussion}--\ref{sec:conclusion}.  We adopt a $h=0.70$,
$\Omega_{\rm M}=0.3$, $\Omega_{\Lambda}=0.7$ cosmology.

\section{Data and Analysis}\label{sec:data}

In Figure~\ref{fig:image}, we show our targeted foreground galaxy
(z=0.19976), producing the observed {\MgII} absorption, and background
quasar (z=0.77) J165931+373528 that are separated by 58{\kpc} projected,
on the sky. Below we describe the data acquired for our analysis.

\subsection{Galaxy and Quasar Spectroscopy}

The galaxy and background quasar spectra were obtained on 2013 April
11 using the Keck Low Resolution Imaging Spectrometer
\citep[LRIS;][]{oke95} with the blue 1200 lines/mm grism blazed at
3400{\A} providing wavelength coverage from the atmospheric cut-off
to 3890{\A}. We further used the red 900 lines/mm grating with a
central wavelength of 7750{\A}. We used a 1$''$ slit and a spectral
binning of two, providing a dispersion of 0.48{\A}~pix$^{-1}$ and
$\sim$1.6{\A} resolution ($\sim$145{\kms}) in the blue and a
dispersion of 1.06{\A}~pix$^{-1}$ and $\sim$3.5{\A} resolution
($\sim$140{\kms}) in the red. The slit was placed such that it
spatially covered both the galaxy and the quasar
(Figure~\ref{fig:image}).  The total of integration time was 2240s.

An additional galaxy spectrum was obtained on 2013 August 12 using the
Double Imaging Spectrograph (DIS) at the Apache Point Observatory
(APO) 3.5-m telescope.  We placed the 1.5$''$ slit along the galaxy
major axis (Figure~\ref{fig:image}) to obtain the kinematic velocity
zeropoint from the {\Ha} emission-line. We used the R1200 grating,
providing a dispersion of 0.58{\A}~pix$^{-1}$ and $\sim$1.3{\A}
resolution ($\sim$50{\kms}).  The wavelength coverage is 1160{\A}
and we used a wavelength centroid of 8100{\A}. The total of
integration time was 4400s.

We supplement our data with a {\it HST}/COS quasar spectrum taken with
the G160M grating (PI:Nestor, PID:12593) in order to measure the
physical properties of the transverse gas derived from {\Lya},
{\SiII}, {\SiIII}, {\MgII} and {\MgI} absorption. The total exposure
time is 2100s and has a wavelength coverage of 1408--1776{\A}.

All the spectra were reduced using standard techniques and were
heliocentric and vacuum corrected. The spectral analysis was performed
using our in-house software. Transverse absorption-line rest-frame
equivalent widths are listed in Table~\ref{tab:abs}.

\subsection{Galaxy Imaging}

In Figure~\ref{fig:image} we show 30$''$$\times$30$''$ section of a
1981s $i$-band image obtained using MegaCam and was produced by the
MegaPipe pipeline \citep{gwyn12}, which has a spatial resolution of
0.186~$''$~pix$^{-1}$.  We used our own custom MCMC bayesian code to
determine galaxy morphological parameters by fitting a S{\'e}rsic
profile convolved with the image point spread function that produced a
S{\'e}rsic index of $n$=0.91$\pm$0.01, a disk effective radius of
$r_e=$1.54$\pm$0.01$''$ and an inclination of
$i=$52$\pm$5~degrees. The quasar is nearly aligned with the galaxy
projected minor axis with a galaxy position angle of
$PA=-3.5\pm0.3$~degrees, which provides ideal geometry to probe
galactic winds \citep{bordoloi11,bouche12,kacprzak12a}. The
galaxy/quasar projected separation is $D=$17.53$''$ (57.8{\kpc}).

\subsection{Galaxy Star Formation Rate}

We estimate lower limits on the unobscured star formation rate (i.e.,
not extinction corrected) using both the H$\alpha$ flux within the
LRIS spectroscopic aperture and the flux density measured from a GALAX
NUV image covering rest-frame wavelengths from 1476 to 2359{\A}. The
SFR of 4.6{\Mpyr} derived from the GALAX UV continuum measurement
following \citet{kennicutt98} provides the more accurate lower bound
on the SFR because a substantial aperture correction must be applied
to obtain the total H$\alpha$ galaxy flux. Alternatively, fitting the
22~$\mu$m flux or the UV-to-17~$\mu$m SED with the CIGALE code
\citep{noll09}, we estimate the dust-corrected SFR could be as large
as 15{\Mpyr} and a stellar mass of log$M_*=10.6\pm0.07$.

\begin{figure}
\begin{center}
\includegraphics[angle=270,scale=0.344]{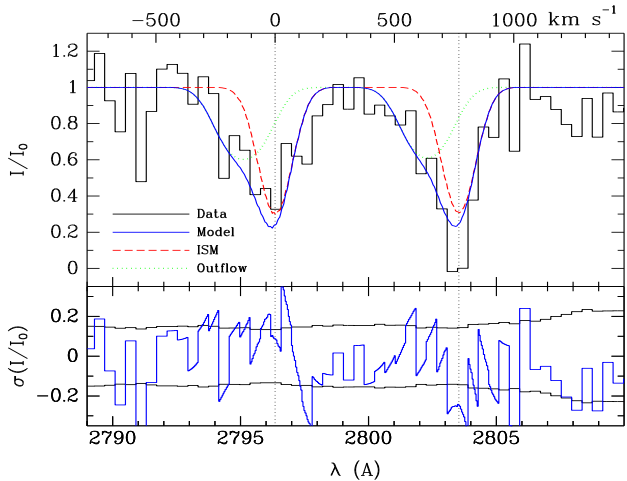}
\caption[angle=0]{The down-the-barrel {\MgIIdblt} doublet absorption
  is fit using two components. One component is fixed to galaxy
  systemic velocity (vertical dotted-line) having a fixed velocity
  width of 200{\kms} (representing the ISM - red dashed line) and
  another component having a variable velocity width representing the
  outflowing gas (green dotted-line). Below is the fit residuals with
  one sigma errors. The outflow absorption component centroid is
  blueshifted by $132\pm25${\kms}. For gas launched perpendicular to
  the disk, this corresponds to a wind velocity of $214\pm41${\kms}.}
\label{fig:martin}
\end{center}
\end{figure}

\section{Results}\label{sec:results}

In Figure~\ref{fig:image}, we show the first known case where {\MgII}
absorption is detected at: (1) intrinsic to the galaxy (i.e.,
``down-the-barrel'') and (2) along the quasar sight-line 58{\kpc} away
(i.e., ``transverse''). The velocity zeropoint is set by the {\Ha}
emission-line centroid, taken along the major axis with DIS, was
determined to be $z=$0.19976$\pm$0.00003. This is consistent with the
minor axis {\Ha} centroid obtained with LRIS. Note that the
down-the-barrel {\MgII} absorption is blueshifted relative to the
galaxy systemic velocity. The down-the-barrel rest-frame {\MgII}
equivalent widths are $W_r(2796)=0.73\pm0.31${\A} and
$W_r(2803)=1.47\pm0.26${\A}.  The observed doublet ratio is less than
unity, for this likely saturated system, which could be due to
optically thin {\MgII} emission from gas within/around the galaxy
\citep{martin12}.

The transverse {\MgII} absorption also exhibits a blueshift relative
to the galaxy systemic velocity. The transverse rest-frame {\MgII}
equivalent widths are $W_r(2796)=0.88\pm0.04${\A} and
$W_r(2803)=0.70\pm0.04${\A}.  Given the relative blueshifts, we
explore whether outflows originating from the galaxy can explain the
kinematics of the transverse absorption at 58{\kpc}.

\subsection{``Down-the-Barrel'' Absorption Wind Model}

The velocity of the {\MgII} absorption is notably blueshifted relative
to the galaxy systemic velocity. A single Gaussian fit to the data
shows that the absorption is blueshifted by $43\pm15${\kms}.  In order
to estimate the bulk velocity of the {\MgII} wind component, we
applied a two-component absorption model. One component is fixed to
systemic velocity of the galaxy with a fixed velocity width of
200{\kms}, as measured from galaxy H$\alpha$ emission-line
(representing the ISM), and one component that has a variable velocity
width that represents the outflowing gas \citep{martin12}. The model
was convolved with a Gaussian profile to model the LRIS instrumental
spread function (ISF). The covering fraction of the variable velocity
component is fixed to unity. The doublet ratio is free parameter and
the residuals between the model and the data are minimized iteratively
using the Levenberg-Marquardt algorithm \citep{press92}.  In
Figure~\ref{fig:martin} we show the best resulting model fit. As
noted, the ISM component of the fit is at the systemic velocity of the
galaxy while the outflow component centroid is blueshifted by
$132\pm25${\kms}.

We note that the velocity offset computed above may not directly
represent the outflow velocity component since the ratio of equivalent
widths is nonphysical, which leads us to suspect the absorption is
partially filled in by resonance {\MgII} emission.  To test this, we
apply models containing an additional emission component.  Since the
data do not resolve an emission component, we attempt to gain insight
by assuming that the emission is at the systemic velocity and its
amplitude is fixed at a modest, but random, value.  Our models show
that the addition of the emission component does not significantly
improve the fit statistic, although we find that including it tends to
lower the inferred outflow velocity.

In summary, our single and double-line models indicate that the
line-of-sight outflow velocity component centroid likely ranges
between $28-157${\kms}. 


\begin{figure*}
\begin{center}
\includegraphics[angle=0,scale=0.29]{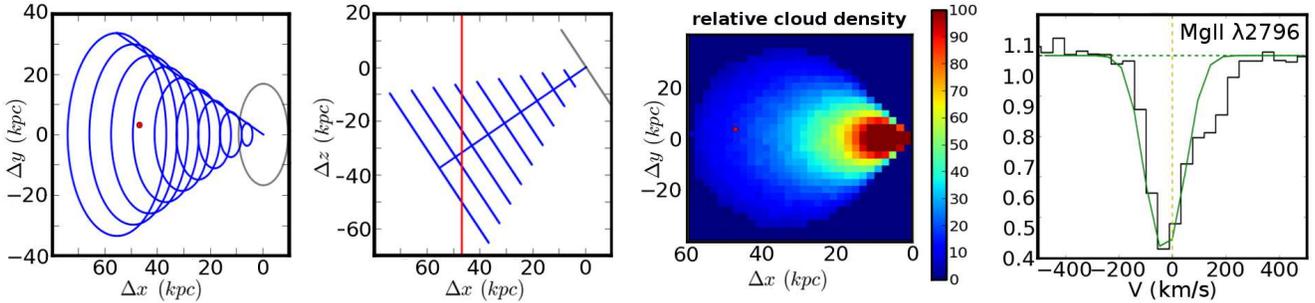}
\caption[angle=0]{(left) Kinematic conical wind model viewed on the
  sky-plane where the $x$-axis corresponds to the minor axis and the
  $y$-axis to the major axis. The grey oval represents an inclined
  galaxy and the circles (blue) represent the conical outflow. The red
  point represents the background quasar. (middle-left) Kinematic
  model shown along the quasar sight-line ($z$-axis) indicated by the
  solid red line. (middle-right) The average cloud relative density
  map with the red point indicating the quasar sight-line. (right) The
  {\MgII$\lambda2796$} transverse absorption (black histogram) with
  the distribution of the cloud line-of-sight velocities (convolved
  with the LRIS resolution) at the quasar location (green). The shape,
  width, and velocity offsets of the synthetic {\MgII} absorption
  constrain the model to have cloud outflow velocities of
  $V_{outflow}=40-80${\kms}.}
\label{fig:winds}
\end{center}
\end{figure*}

\begin{figure}
\begin{center}
\includegraphics[angle=0,scale=0.95]{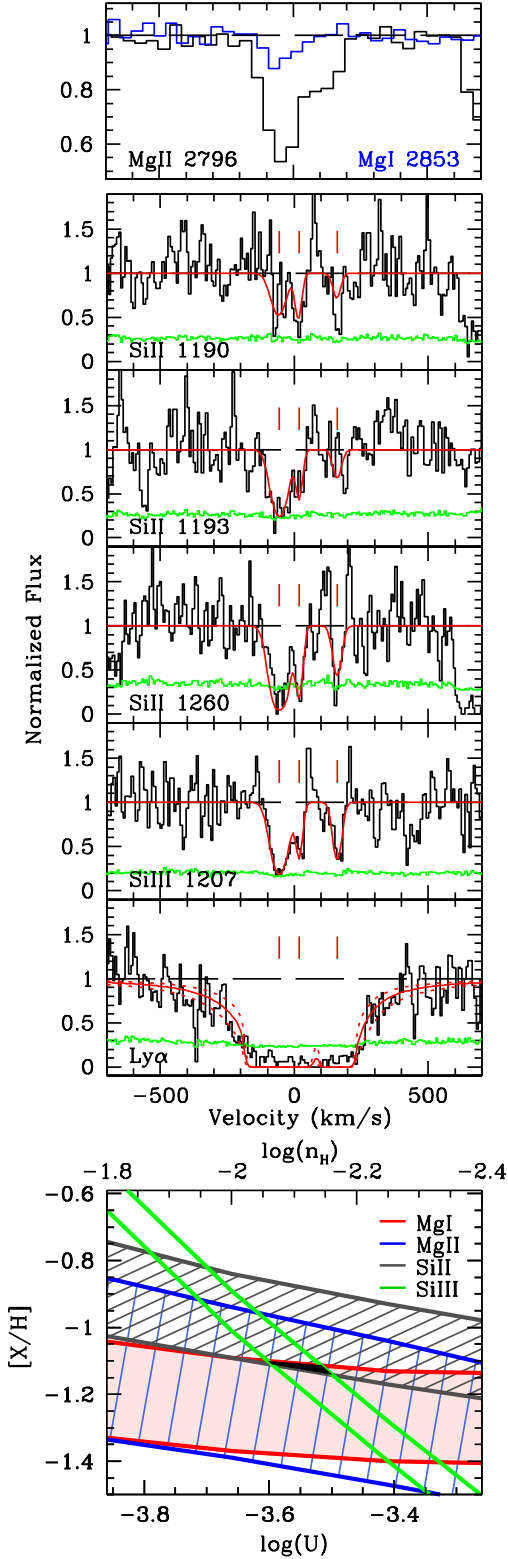}
\caption[angle=0]{ (top) Transverse {\MgII}$\lambda$2796 (black) and
  {\MgI}$\lambda$2853 (blue) absorption. (middle) The {\it HST}/COS
  spectrum of the transverse {\SiII}, {\SiIII} and Ly$\alpha$
  absorption. The data are shown as a histogram (black) and the sigma
  spectrum shown below (green). The data have been Voigt profile fit
  (red), with the number of clouds indicated by the vertical
  tick-marks. The N({\HI})$=18.89\pm0.15$ with the errors on the fit
  shown as dotted lines. --- (bottom) The Cloudy models are shown as a
  function of ionization parameter and metallicity. The upper and
  lower bounds for each ion include the N(${\rm X}$) and the N({\HI})
  errors. The overlapping/common region, indicated by the black shaded
  region, is where the data constrain the acceptable range of
  metallicities to be $-1.14\leq [{\rm X/H}]\leq-1.10$ and ionization
  parameter to be $-3.60\leq U\leq-3.50$.}
\label{fig:cos}
\end{center}
\end{figure}
\begin{center}
\begin{deluxetable}{llrr}
\tabletypesize{\scriptsize} \tablecaption{Transverse Absorption
  Properties\label{tab:abs}} \tablecolumns{4} \tablewidth{0pt} 

\tablehead{
\colhead{Ion}&
\colhead{Telescope/}&
\colhead{$W_r$} &
\colhead{log[$N({\rm X})$]} \\
\colhead{} &
\colhead{Instrument} &
\colhead{(\AA)}&
\colhead{ } 
}
\startdata
Ly$\alpha$             & {\it HST}/COS  &2.03$\pm$0.07  &      \\
{\HI}                  &                &               &    18.89$\pm$0.15\\[1.0ex]
{\MgII} $\lambda$2796  & Keck/LRIS      &0.88$\pm$0.04 &          \\
{\MgII} $\lambda$2803  & Keck/LRIS      &0.70$\pm$0.04  &         \\
{\MgII}                &                &               &    13.62$\pm$0.18\\[1.0ex]
{\MgI}$\lambda$2853    & Keck/LRIS      &0.13$\pm$0.04  &          \\
{\MgI}                 &                &               &    12.05$\pm$0.13\\[1.0ex]
{\SiII} $\lambda$1190  & {\it HST}/COS  &0.33$\pm$0.05  &          \\
{\SiII} $\lambda$1193  & {\it HST}/COS  &0.37$\pm$0.04  &          \\
{\SiII} $\lambda$1260  & {\it HST}/COS  &0.44$\pm$0.06  &         \\
{\SiII}                &                &               &    13.93$\pm$0.07\\[1.0ex]
{\SiIII} $\lambda$1207 & {\it HST}/COS  &0.42$\pm$0.05  &          \\
{\SiIII}               &                &               &    13.54$\pm$0.06\\[1.0ex]
{\SiIV} $\lambda$1394  & {\it HST}/COS  &$<$0.62        &     \\
{\SiIV} $\lambda$1403  & {\it HST}/COS  &$<$0.59        &        \\  
{\SiIV}                &                &               &       $<$12.80
\enddata
\end{deluxetable}  
\end{center}

\subsection{Transverse Absorption Wind Model}

Here we explore whether or not the {\MgII} absorption detected at
58{\kpc} is consistent with an outflow model. We adopt a bi-conical
outflow \citep{bouche12} since the gas trajectories will tend to open
up from the internal wind pressure into a cone-like shape
\citep{shen12}. The model has two free parameters, the cone
half-opening angle ($\theta_{max}$) and the cloud velocities, which
are radial and are assumed to be constant with radius ($V_{outflow}$).
These are the only two free parameters since the relative geometric
orientation of the wind with respect to the quasar line-of-sight is
provided by the deep CFHT galaxy image. We allow $\theta_{max}$ to
range from 30--45 degrees, which is consistent with previous
observations \citep{bouche12,kacprzak12a}. The observed blueshifted
{\Ha}--{\MgII} relative velocity offset constrains that the modeled
outflow cone is pointing towards the observer.

In Figure~\ref{fig:winds}, we show a 48 degree inclined cone with
$\theta_{max}=45$~degrees in the plane of the sky. We adopt the
convention that the $x$ and $y$ axes represent the plane of the sky
and are aligned along the galaxy minor and major axes, respectively,
while the quasar sight-line is orthogonal to the sky-plane ($z$-axis).
The grey oval represents the inclined disk and the blue circles
represent the conical outflow. The quasar location is shown in red.
We show the cloud $z$-velocities as a function of position. We also
show the line-of-sight velocity distribution of the clouds at the
location of the quasar, which closely mimics the shape and width of
the transverse {\MgII} absorption data. This distribution is convolved
with an instrumental resolution of 145{\kms}, similar to the LRIS
data.

We constrained the model wind velocity by enforcing agreement between
the wind model {\MgII} absorption profile shape, width, and velocity
offset and the observed transverse {\MgII} absorption profile (see
Figure 3).  The range of radial model winds speeds that provides the
best $\chi^2$ fit to the data are $V_{outflow}=40-80${\kms}.

In order to directly compare the bi-conical wind model radial
velocities to the measured line-of-sight down-the-barrel velocities,
we must deproject the down-the-barrel velocities by the galaxy
inclination, which provides the perpendicular velocity component
emanating from the galaxy disk (which is equivalent to the radial
velocity of the axis of symmetry of the wind model). The range of
deprojected down-the-barrel velocities is $V_{dtb}=45-255${\kms},
which is consistent with, though possibly faster than, the
$V_{outflow}=40-80${\kms} deduced from the bi-conical wind model.

We note that we are unable to reproduce the reddest {\MgII} component
(also seen in silicon), which could be due to stochastic effects from
the number of clouds intercepted along the line-of-sight or to
limitations of our constant wind velocity, geometrically symmetric
model, and/or the absorption may arise from other sources within the
galaxy halo.  Our model is not the only possible explanation for the
absorption and without a more complex model, it is difficult to
conclude the origins of the reddest component, however, our simple
model does successfully reproduce the majority of the {\MgII} and all
of the {\MgI} absorption.

\subsection{Galaxy and Halo Gas Metallicities}

We compute the galaxy metallicity using the indicator
$N2=$log$[f([$\NII$]\lambda6583)/f($H$\alpha)]$, which is equivalent
to being the ratio of equivalent widths since both H$\alpha$ and
{[\NII]} are only $20.66${\A} apart (at rest wavelengths) and the
continuum flux levels are approximately equal. The rest-frame
equivalent widths, measured from APO/DIS, of {\Ha}=58$\pm$2{\A} and
[\NII]$\lambda6583$=$13\pm1${\A} (shown in Figure~\ref{fig:image})
yield $N2=$$-0.65$$\pm$0.04. We apply the $N2$ metallicity relation
12$+$log[O/H]$=8.90+0.57\times N2$ \citep{pettini04}, assuming a solar
oxygen abundance of 12$+$log[O/H]$_{\odot}=8.736\pm0.078$
\citep{holweger01}, and compute a galaxy metallicity of [{\rm
    O/H}]=$-0.21$$\pm$0.08.

To determine the transverse absorbing-gas properties we Voigt profile
fit the {\it HST}/COS spectrum of {\SiII}, {\SiIII}, {\SiIV} and
{\Lya} using our software MINFIT \citep{cv01,cvc03}, which
incorporates the appropriate COS ISF. In Figure~\ref{fig:cos}, we show
the {\SiII}, {\SiIII} and {\Lya} ({\SiIV} is not shown since it is a
upper limit) along with the Voigt profile fits. The total column densities
are shown in Table~\ref{tab:abs}. The velocity structure of {\Lya} was
established from silicon assuming thermal broadening. We find a
log$N({\HI})$=18.89$\pm$0.15.

The {\MgII} column density was determined by fitting Gaussians to the
{\MgII} profiles using the velocity structure of the {\SiII} lines
from their Voigt profile fits.  We applied the curve-of-growth to each
{\MgII} Gaussian component using the equivalent widths from the
Gaussians and the {\SiII} Doppler $b$ parameters.  For the {\MgI},
only a single Gaussian fit was statistically required. Given small
$W_r$({\MgI}), placing it on the linear part of the curve-of-growth,
the column density is independent of $b$.  The total column densities
are presented in Table~\ref{tab:abs}.

We use version 08.00 of Cloudy \citep{ferland13}, and a solar
abundance pattern, to model the metallicity and ionization conditions
of the gas. We apply the standard assumption that the gas is
represented by a photoionized uniform slab in ionization equilibrium
illuminated with a \citet{haardt12} ionizing spectrum.  The ionization
parameters, $U$, and the metallicity of the gas are varied to match
the observations of $N({\rm X})$ shown in Table~\ref{tab:abs}.  In
Figure~\ref{fig:cos}, we show the Cloudy models as a function of
ionization parameter and metallicity.

The upper range of {\MgI} and lower range of {\SiII} enforce $\log
U>-3.60$.  The steep dependence of {\SiIII}, for which the Voigt
profile fit is very robust, further constrains $\log U<-3.50$.  Thus,
the allowed range of ionization parameter is $-3.60\leq\log
U\leq-3.50$, the range of hydrogen density is $-2.16\leq
n_H\leq-2.06$, the metallicity\footnote{Additional errors from
  model-dependant assumptions could range between $0.1-0.4$dex
  \citep{werk14}.} is $-1.14\leq[{\rm X/H}]\leq-1.10$, the hydrogen
ionization fraction is $-0.77\leq f($\HI$)\leq-0.69$, and the total
hydrogen column density is $19.57\leq$log[N({\rm H})]$\leq19.68$.

\section{Discussion}\label{sec:discussion}

The kinematic models of both the down-the-barrel and transverse
absorption produce consistent predictions for the outflow velocities
that overlap in the range of $45-80${\kms}.  These results are
highly suggestive that the transverse absorption in the quasar
spectrum is physically related to the down-the-barrel absorption via
an outflow.

We have computed the galaxy metallicity to be near solar while the
transverse absorption at 58{\kpc} is a tenth solar. The metallicity of
the transverse component is consistent with the 0.1--1$Z_{\odot}$
metallicity systems around $\sim0.1L_*$ galaxies that have velocities
consistent with being bound, galactic fountain clouds
\citep{stocke13}. The absorption-line metallicity is also consistent
with the higher metallicity outflowing gas from the bi-modal
distribution of Lyman-limit system metallicities from
\citet{lehner13}.

To determine the gas outflow rate we apply Equation~4 of
\citet{bouche12}, derived for a single outflow cone, using our derived
model outflow rates and derived N({\rm H}). We estimate the total
outflow rate from a bi-conical flow being roughly 1.6--4.2{\Mpyr}
assuming outflow velocities of $40-80${\kms}. With the galaxy having a
SFR between $4.6-15${\Mpyr}, we conclude that the wind mass-loading
factor likely lies in the range of $\eta\sim0.1-0.9$, which is
consistent with loading factors derived for star-forming galaxies at
low redshift \citep{martin99,rupke05b,bouche12} but slightly lower
than recent estimates from scattered {\MgII} emission at intermediate
redshifts \citep{martin13}.

\section{Conclusions}\label{sec:conclusion}

We have shown the first example of a galaxy that exhibits absorption
observed down-the-barrel {\it and} transversely at a projected
distance of 58{\kpc}.  Both the {\MgII} observed down-the-barrel and
transversely are blueshifted with respect to the galaxy systemic
velocity. We also detect {\Lya}, {\SiII}, {\SiIII}, and {\MgI}
absorption at the transverse location. The quasar sight-line resides
within $3.5^{\circ}$ of the projected galaxy minor axis where studies
suggest that the absorption should be produced by winds.
\begin{enumerate}

\item We find that the down-the-barrel deprojected outflow velocities,
  i.e., perpendicular to the galaxy disk, range between
  $V_{dtb}=45-255${\kms}, which is typical of star-forming galaxies
  \citep[e.g.,][]{martin12,rubin13}.

\item If we assume a conical outflow model \citep{bouche12}, then the
  constant wind velocities required to reproduce the transverse
  {\MgII} absorption kinematics are $V_{outflow}=40-80${\kms}, which
  is consistent with the deprojected down-the-barrel outflow
  velocities. Although this is a simplistic wind model, our analysis
  suggests that the absorption is kinematically coupled.

\item We compute the galaxy metallicity to be [{\rm
    O/H}]=$-0.21$$\pm$0.08, whereas the transverse absorption at
  58{\kpc} has $[{\rm X/H}]=-1.12\pm0.02$.

\item The galaxy SFR ranges from 4.6 to 15{\Mpyr} while
  the estimated outflow rate is roughly 1.6--4.2{\Mpyr}
  and yields a wind loading factor of $\eta=0.1-0.9$.

\end{enumerate}
For the first time, we have successfully combined independent models
and analysis techniques of down-the-barrel and transverse absorption
systems to show that the intrinsic galaxy outflows sufficient to
reproduce the observed kinematics of the transverse absorption
58{\kpc} away. If the metallicty at the base of the outflow equals
that of the galaxy ISM, and the wind is continuous, then the observed
$\sim$1~dex decrease in metallicity at 58{\kpc} suggests that the gas
was diluted/mixed with lower metallicty gas.  Finding additional
systems like this one will aide in our understanding of how outflows
transport and redistribute gas within their halos.


\acknowledgments  {\ack}





{\it Facilities:} \facility{Keck I (LRIS)}, \facility{Sloan (SDSS)},
\facility{HST (COS)}, \facility{APO (DIS)}.

\end{document}